\documentclass[aps,pra,amsmath,amssymb,notitlepage,nofootinbib,twocolumn]{revtex4-2}
\usepackage{mathtools,xparse}
\usepackage[utf8]{inputenc}
\usepackage{graphicx,hyperref,upgreek,bbm,mdframed,siunitx}
\usepackage{xcolor}
\hypersetup{    colorlinks,    linkcolor={blue!50!black},    citecolor={blue!50!black},    urlcolor={blue!80!black}}
\usepackage[all]{hypcap}
\usepackage{bbold}
\graphicspath{{img/}}
\newcommand*{\bra}[1]{\ensuremath{\langle #1 \vert}}
\newcommand*{\ket}[1]{\ensuremath{\vert #1 \rangle}}

\renewcommand{\v}[1]{\textbf{#1}}

\begin{document}

\title{Neural quantum states for emitter dynamics in waveguide QED}

\author{Tatiana Vovk$^{1,2}$}
\email{tatiana.vovk@uibk.ac.at}
\author{Anka Van de Walle$^{3,4,5}$}
\author{Hannes Pichler$^{1,2}$}
\author{Annabelle Bohrdt$^{3,4}$}

\affiliation{$^1$Institute for Theoretical Physics, University of Innsbruck, 6020 Innsbruck, Austria}
\affiliation{$^2$Institute for Quantum Optics and Quantum Information, Austrian Academy of Sciences, 6020 Innsbruck, Austria}
\affiliation{$^3$Ludwig Maximilian University Munich, 80333 Munich, Germany}
\affiliation{$^4$Munich Center for Quantum Science and Technology, 80799 Munich, Germany}
\affiliation{$^5$Department of Physics and Astronomy, Ghent University, 9000 Ghent, Belgium}

\date{\today}

\begin{abstract}
Quantum emitters coupled to one-dimensional waveguides constitute a paradigmatic quantum-optical platform for exploring collective phenomena in open quantum many-body systems. For appropriately spaced emitters, they realize the Dicke model, whose characteristic permutation symmetry allows for efficient exact solutions featuring superradiance. When the emitters are arbitrarily spaced, however, this symmetry is lost and general analytical solutions are no longer available. In this work, we introduce a novel numerical method to study the dynamics of such systems by extending the time-dependent neural quantum state (t-NQS) framework to open quantum systems. We benchmark our approach across a range of waveguide QED settings and compare its performance with tensor-network calculations. Our results demonstrate that the t-NQS approach is competitive with other numerical methods and highlight the potential of t-NQSs for studying open quantum many-body systems out of equilibrium.
\end{abstract}
\maketitle

\section{Introduction}
Waveguide quantum electrodynamics (QED) platforms are a class of open quantum many-body systems, where multiple quantum emitters (such as qubits) interact through a shared photonic environment~\cite{kimble2008quantum, pichler_quantum_2015, chang_colloquium_2018, turschmann2019coherent, sheremet_waveguide_2023}. In these systems, emitters are coupled identically to a one-dimensional bath, giving rise to photon-mediated long-range interactions~\cite{shahmoon2013nonradiative, pichler_quantum_2015, mok_dicke_2023}. Such setups enable the exploration of rich many-body dynamics~\cite{masson_many-body_2020, fayard_many-body_2021, cardenas-lopez_many-body_2023, mahmoodian_dynamics_2024, poshakinskiy_quantum_2024, wang_boundary_2025} and serve as platforms for practical applications, including dissipative state preparation~\cite{agusti2023autonomous, agarwal2024directional, rubies-bigorda_deterministic_2024}, quantum memory~\cite{gorshkov2007universal, gouraud2015demonstration, sayrin2015storage} and quantum information processing~\cite{royer2018itinerant, grimsmo2021quantum, agusti2023autonomous}. Experimental realizations of waveguide QED systems span a broad range of physical systems, including cold atoms near nanophotonic waveguides~\cite{goban2014atom, corzo2019waveguide}, superconducting qubits coupled to microwave transmission lines~\cite{mirhosseini2019cavity, kannan2020waveguide}, and solid-state emitters such as color centers~\cite{sipahigil2016integrated} and quantum dots~\cite{foster2019tunable} coupled to nanofibers.

Although an extensive range of waveguide QED and quantum spin network models have been studied~\cite{cirac1997quantum, chang2007single, gonzalez2013mesoscopic, pichler_quantum_2015, kumlin2018emergent, mann2024selective, agarwal2024directional, ilin_correlated_2024}, much of the foundational theoretical and numerical work has been built around the paradigmatic Dicke model~\cite{dicke_coherence_1954, leppenen_quantum_2024, holzinger_solving_2025}. In this model, an ensemble of $n$ excited qubits emits light collectively, which is referred to as \textit{Dicke superradiance}~\cite{gross_superradiance_1982}.
The Dicke model implies full permutational symmetry among emitters, which allows for a dramatic reduction of the effective Hilbert space. Specifically, an ensemble of $n$ qubits can be described as a single collective spin-$n/2$, with a Hilbert space dimension of $\left(n+1\right)$~\cite{lee_exact_2025}. This symmetry facilitates analytical treatment and enables efficient numerical simulations~\cite{bassler2025absence, zhang2025unraveling, rosario2025unraveling, holzinger_solving_2025}.

It is of interest to understand how the collective effects including superradiance are modified once this idealized symmetry is broken, which is inevitable in realistic setups. In waveguide QED, for instance, this occurs when the emitters are displaced from positions commensurate with the wavelength corresponding to the transition frequency. In such cases, the short-time dynamics can still be computed efficiently~\cite{lee_exact_2025, holzinger_beyond_2025}. However, since the system’s state is no longer permutationally symmetric, determining the transient and long‑time dynamics becomes substantially challenging, as one typically needs to work in the full Hilbert space, which scales exponentially with $n$. This requires development of novel numerical methods capable of capturing complex open quantum many-body dynamics.

Among the emerging numerical methods for complex quantum many-body systems, neural quantum states (NQSs) have recently gained significant attention~\cite{carrasquilla_neural_2021, lange_architectures_2024, schmitt_simulating_2025}. In this approach, neural networks (NNs) are employed to represent quantum many-body states, offering potentially greater expressibility than conventional tensor network (TN) techniques~\cite{carleo_solving_2017}. In the domain of open quantum systems, NQSs have been successfully applied to problems involving \textit{local} noise, particularly for steady-state computations~\cite{vicentini_variational_2019, yoshioka_constructing_2019, nagy_variational_2019, hartmann_neural-network_2019, vicentini2022positive} and the simulation of dissipative dynamics~\cite{hartmann_neural-network_2019, reh_time-dependent_2021, luo_autoregressive_2022, vicentini2022positive}. In contrast, collective phenomena in waveguide QED setups typically involve \textit{global} quantum noise. Within the NQS framework, such global noise channels have so far been addressed only in the context of steady-state analysis~\cite{wei_variational_2025}. Here, we extend the NQS framework to capture the full real-time dynamics as well.

In this work, we present a novel NQS-based numerical method for simulating the transient dynamics of quantum spin networks with arbitrary arrangement of qubits coupled to a Markovian one-dimensional (1D) waveguide. We employ the recently introduced time-dependent NQS (t-NQS) approach~\cite{de_walle_many-body_2024}, extending it to the case of open-system dynamics. We demonstrate that our method captures the dynamics in the generic, symmetry-broken regime {beyond the standard Dicke model}: in the case of incommensurate emitter spacings, it exhibits improved convergence compared to the simulations accessible by the conventional TN method. We also demonstrate the versatility of physical regimes in waveguide QED that can be simulated using our method. Finally, we apply the developed approach to investigate how superradiance is affected by a specific symmetry breaking that shifts the system away from the Dicke model, confirming a quadratic decrease in the peak emission intensity as the distance between emitters increases.

This paper is organized as follows. In Section~\ref{sec:phys_desc}, we introduce the notation and give the description of the waveguide QED setup studied in this work. In Section~\ref{sec:TN}, we {analyze the performance of standard TN methods when applied to} systems without permutational symmetry. Section~\ref{sec:NQS} introduces our t-NQS-based method for solving the dynamics of open quantum many-body systems. In Section~\ref{sec:results}, we benchmark both approaches across several physical regimes, and then use the t-NQS method to calculate emitter dynamics in generic waveguide QED settings.

\begin{figure}
\label{fig:1}
    \centering
    \includegraphics[width=0.4\textwidth]{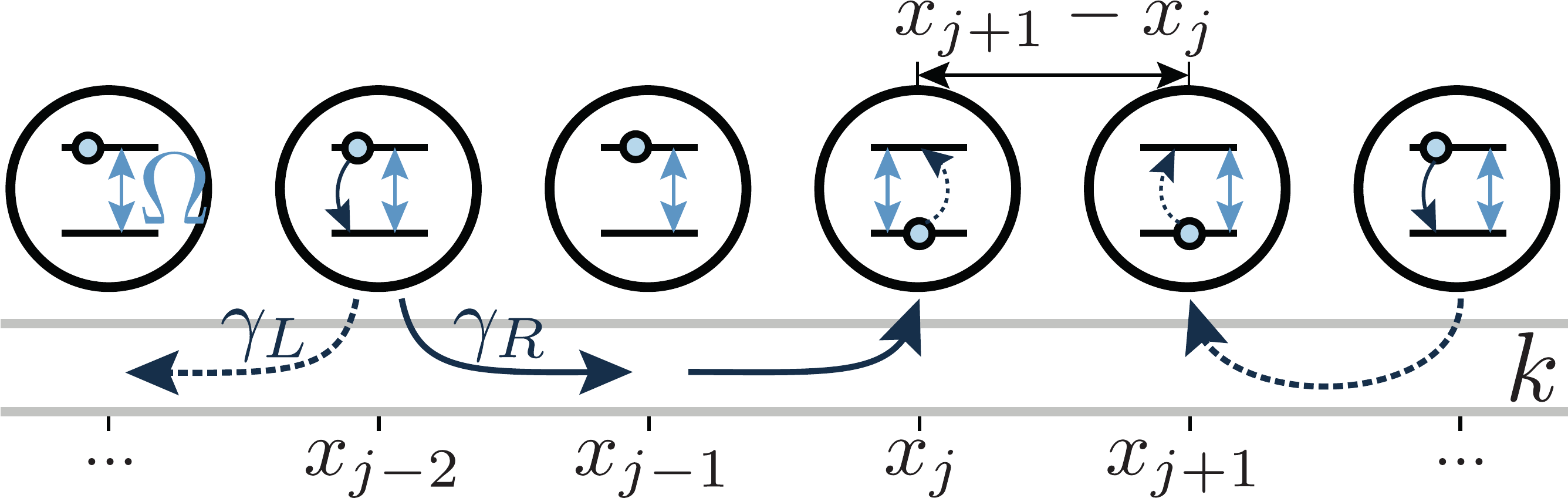}
    \caption{Chain of spin-$1/2$ emitters located at positions $x_j$ near a one-dimensional Markovian waveguide with wavenumber $k$ and decay rates into the left- and right-propagating photonic modes, denoted by $\gamma_L$ and $\gamma_R$, respectively. The emitter chain can be additionally globally driven with a Rabi frequency $\Omega$.}
\end{figure}

\section{Physical description}
\label{sec:phys_desc}
In this section, we briefly describe the waveguide QED model under investigation. Consider a {system} that consists of a chain of $n$ emitters, each with a local Hilbert space dimension $d = 2$ (also referred to as qubits or spin-$1/2$ particles), trapped near a one-dimensional, Markovian, open-ended waveguide that acts as a shared {environment}, as illustrated in Fig.~\ref{fig:1}. Emitters can undergo a radiative transition from the excited state to the ground state by emitting a photon into the left- and right-propagating modes of the waveguide with rates $\gamma_L$ and $\gamma_R$, respectively. Throughout the text, we denote the two states of the $j^\mathrm{th}$ spin as $\ket{0}_j$ and $\ket{1}_j$, corresponding to spin-down and spin-up, respectively. We are interested in the dynamics of a spin chain that is initialized in the all-spin-up state:
\begin{align}
    \rho_0 = \bigotimes_{j=1}^n\ket{1}_j\bra{1}.
    \label{eq:init}
\end{align}
The spin system obeys the following Lindblad master equation (ME)~\cite{lindblad_completely_1975, pichler_quantum_2015, mok_universal_2024}:
\begin{align}
    \partial_t{\rho} = -i \left[H +H_\mathrm{diss}, \rho\right] + \mathcal D\left[\rho\right],
    \label{eq:ME}
\end{align}
where $H = \Omega \sum_j \sigma_j^x$ is the system Hamiltonian corresponding to a local driving with Rabi frequency $\Omega$, $H_\mathrm{diss}$ is the coherent term induced by the dissipation and $\mathcal D\left[\rho\right]$ is the dissipation superoperator~\cite{pichler_quantum_2015}. We call the ME~\eqref{eq:ME} \textit{balanced}, when {the emission rates into the left- and right-moving bath modes shown in Fig.~\ref{fig:1} are the same}, $\gamma_L = \gamma_R \equiv \gamma$. In this case, the terms $H_\mathrm{diss}$ and $\mathcal D\left[\rho\right]$ take the form\footnote{We refer the reader to Appendix~\ref{app:ME_gen} for the general form of the ME in the case when the waveguide is \textit{imbalanced}, $\gamma_L \neq \gamma_R$.}:
\begin{subequations}
\label{eq:H_D_balanced}
\begin{align}
    &H_\mathrm{diss} = {\gamma} \sum_{j,\ell} \sin \nu_{j\ell} \sigma_j^+ \sigma_\ell^-, \label{eq:H_diss_balanced} \\
    &\mathcal D\left[\rho\right] = {2 {\gamma}} \sum_{j,\ell} \cos\nu_{j\ell} \left(\sigma_\ell^- \rho  \sigma_j^+ - \frac{1}{2} \left\{\sigma_j^+ \sigma_\ell^-, \rho\right\}\right), \label{eq:D_balanced} \\
    &\text{with}~\nu_{j\ell} = k \delta x_{j\ell} \equiv k \left|x_j - x_\ell\right|,\nonumber
\end{align}
\end{subequations}
where $k$ is the wavenumber of the guided mode, $x_j$ are the spin positions along the waveguide, $\sigma_j^+ = \ket{1}_j \bra{0}$ and $\sigma_j^- = \ket{0}_j \bra{1}$ are the $j^\mathrm{th}$ spin operators. We note that the decoherence rates are rescaled with the system size, $\gamma = \Gamma / n$, and we use the constant $\Gamma$ as a base time scale throughout this work~\cite{carollo_exact_2022, cabot_quantum_2023}. Eq.~\eqref{eq:ME} generally involves all-to-all two-body interactions and leads to a complicated intermediate dynamics and, in case of zero driving, $\Omega = 0$, a trivial steady state, $\rho_\mathrm{SS} = \bigotimes_{j=1}^n\ket{0}_j \bra{0}$~\cite{mirza2016multiqubit}.

When the distances between the emitters (or emitter \textit{spacings}) are commensurate with the wavelength corresponding to the transition frequency, such that $k \delta x_{j\ell} = 2\pi m$, $\forall j,\ell$ with $m \in \mathbb{Z}^+$, the coherent part~\eqref{eq:H_diss_balanced} of the balanced ME vanishes and the system reduces to
a permutationally symmetric Dicke model~\cite{dicke_coherence_1954, gross_superradiance_1982}:
\begin{align}
    \partial_t{\rho} = -i \left[H, \rho\right] + {2{\gamma}} \sum_{j,\ell} \left(  \sigma_\ell^- \rho  \sigma_j^+ - \frac{1}{2} \left\{\sigma_j^+ \sigma_\ell^-, \rho\right\}\right).
    \label{eq:Dicke}
\end{align}
When the system is initialized as~\eqref{eq:init} and evolves under~\eqref{eq:Dicke}, its dynamics can be reduced to that of a collective spin-$n/2$ and can be solved exactly in an efficient way~\cite{bassler2025absence, zhang2025unraveling, rosario2025unraveling, holzinger_solving_2025}. In this setting, when $H=0$, the system exhibits the phenomenon of \textit{Dicke superradiance}, when the spins emit photons collectively {with an intensity that is larger by a factor of $n^2$ than that of an independent emitter}~\cite{mok_dicke_2023}.

When however the emitter spacings are \textit{incommensurate} with $k$, the permutational symmetry is broken and the dynamics can no longer be reduced to that of a single large spin. In this case, the many-body ME~\eqref{eq:ME} has to be solved explicitly in the unconstrained Hilbert space. {The exact diagonalization (ED) methods in this case are limited to small system sizes due to the exponential increase of the Hilbert space}. In the following section we demonstrate an attempt to simulate the incommensurate chain dynamics with the standard tensor network (TN) technique based on the vectorization of the ME~\eqref{eq:ME}~\cite{verstraete2004matrix} and show that with this method it is difficult to converge the simulations for larger system sizes.

\section{Challenges of tensor network solutions}
\label{sec:TN}
In this section we present a TN-based approach to solve Eq.~\eqref{eq:ME} and limitations of this type of solutions. We first provide the reader with the basic considerations concerning the implementation of the matrix product state (MPS) solution and then show how, once deviating from the standard Dicke model~\eqref{eq:Dicke}, it becomes hard for the TN approach to represent the exact state dynamics.

The standard TN technique for open quantum many-body systems involves vectorizing the density matrix, $\rho = \sum_{\alpha,\beta} \rho_{\alpha\beta} \ket{\alpha} \bra{\beta} \rightarrow \ket{\mathbf{\rho}}\rangle = \sum_{\alpha,\beta} \rho_{\alpha\beta} \ket{\alpha} \otimes \ket{\beta} $, and representing it as an MPS in the Liouville (double Hilbert) space~\cite{verstraete2004matrix, schollwock_density_2011}. In this case, the operator of the state propagation under the ME~\eqref{eq:ME} can be efficiently encoded into a matrix product operator (MPO) using a {finite automata} approach~\cite{crosswhite2008finite, hubig2017generic}. We refer the reader to Appendix~\ref{app:MPO_prop} for an explicit form of the propagator. Importantly, the propagator MPO has a fixed bond dimension $\chi_\mathrm{prop} = 10$ for any system size.

For a given MPS, the computational cost of the TN processing $\mathcal C_\mathrm{TN}$ is determined by its \textit{bond dimension} $\chi$ as well as the system size $n$ (see also Appendix~\ref{app:costs}):
\begin{align}
    \mathcal C_\mathrm{TN} \left(\chi, n\right)= n \chi^3.
    \label{eq:MPO_cost}
\end{align}
An associated operator entanglement (OE) entropy acts as a proxy of this cost~\cite{wellnitz2022rise, preisser_comparing_2023}:
\begin{align}
    E = - \sum_{i=1}^\chi \mu^2_i \log_2 \mu^2_i,~~\mathrm{with~} \mu_i = \frac{\tilde \mu_i}{\sqrt{\sum_{i'} \tilde\mu^2_{i'}}},
    \label{eq:OE}
\end{align}
where $\tilde \mu_i$ ($\mu_i$) are the (normalized) singular values of the MPS’s middle cut. Depending on the amount of correlations present in the state $\ket{\rho}\rangle$, its full TN representation may require different bond dimensions $\chi$, and thus different levels of computational effort. This difference can be quantified by analyzing how the bond dimension $\chi$ and the corresponding OE~\eqref{eq:OE} scale with the system size $n$~\cite{eisert2010colloquium}. For instance, if the OE scales linearly with $n$, the corresponding state exhibits a \textit{volume‑law} scaling of entanglement, and the exact representation requires the bond dimension to grow exponentially with $n$, matching the scaling of a naive full‑state representation. On the contrary, when the OE saturates at a constant value (an \textit{area‑law} scaling), the bond dimension grows linearly, allowing an efficient state representation. An intermediate situation implies bond dimension growth that is \textit{polynomial} in $n$.

\begin{figure}
\label{fig:2}
    \centering
    \includegraphics[width=0.45\textwidth]{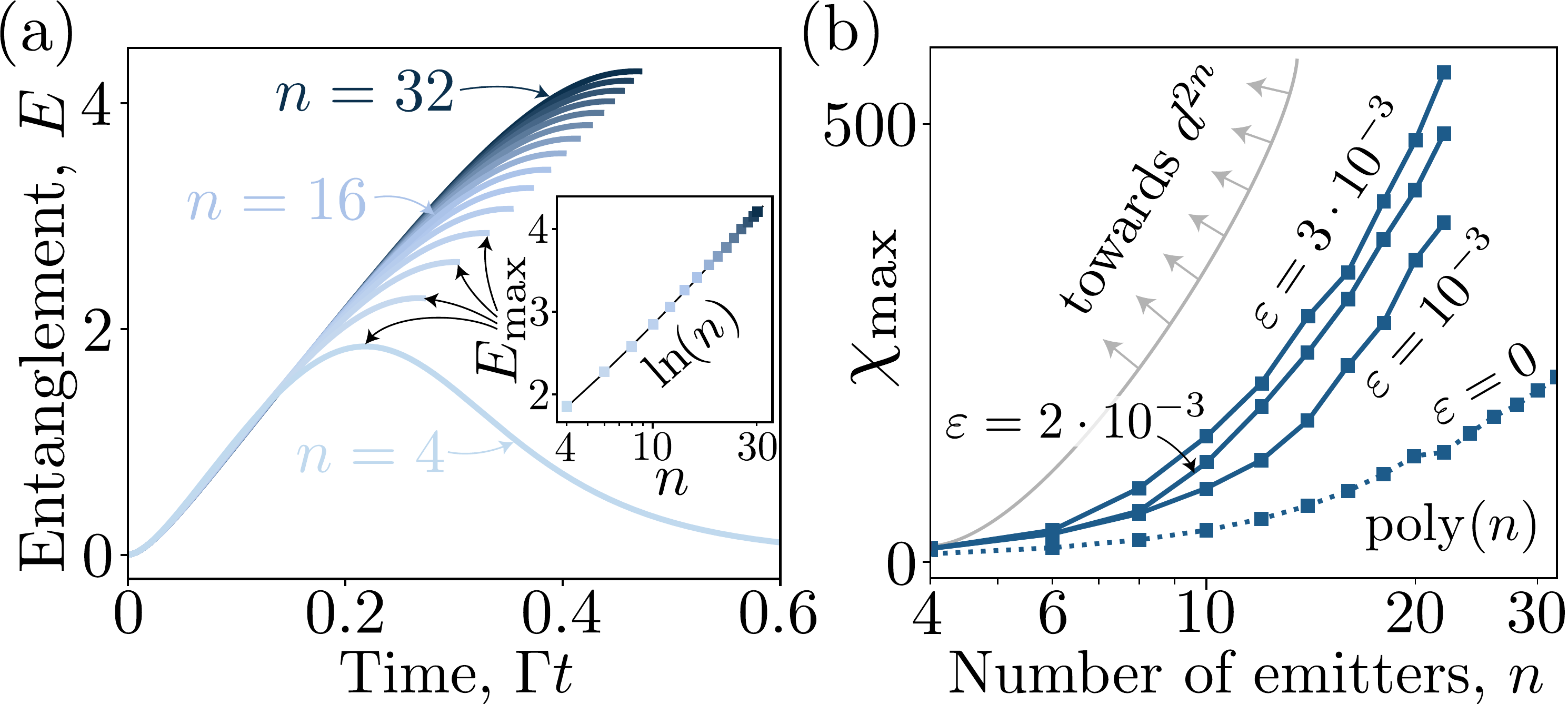}
    \caption{Characteristics of tensor network (TN) solutions for the long-range waveguide QED problem. (a) Operator entanglement (OE) dynamics in the Dicke regime for various system sizes ranging from $n = 4$ to $n = 32$ (indicated by different colors). Inset: scaling of the maximum OE with system size $n$. (b) Scaling of the maximum bond dimension $\chi_\mathrm{max}$ with system size, for emitter spacings parametrized as in Eq.~\eqref{eq:incomm_spacing}.
    }
\end{figure}

To asses the computational cost of solving the dynamics of $\rho$ under the ME, we consider the scaling of the maximal OE encountered during the dynamics $E_\mathrm{max}$ with the system size $n$. In Fig.~\ref{fig:2}(a) we solve the Dicke ME~\eqref{eq:Dicke} (commensurate emitter spacings, $k \delta x_{j,j+1} = 2\pi$). We start from the state~\eqref{eq:init} and show evolutions of the OE for different system sizes. One can observe a rapid entanglement buildup up to a maximal value $E_\mathrm{max}$ with a subsequent relaxation towards zero entanglement in the steady state, $E\left(\rho_\mathrm{SS}\right) = 0$. This phenomenon is referred to as the \textit{entanglement barrier} and is typical for open quantum systems~\cite{wellnitz2022rise, preisser_comparing_2023}. In the inset of Fig.~\ref{fig:2}(a) we confirm the logarithmic scaling of $E_\mathrm{max}$ with the system size $n$ in the Dicke regime. This corresponds to the polynomial scaling of the maximal bond dimension $\chi_\mathrm{max}$ required to represent the state with the OE $E_\mathrm{max}$, which is shown as a dashed line in Fig.~\ref{fig:2}(b).

It turns out that the entanglement dynamics significantly depends on the emitter positions $x_j$, even though the MPO propagator's size does not depend on this (see Appendix~\ref{app:MPO_prop}). Once the emitter spacings are slightly incommensurate:
\begin{align}
    k \delta x_{j,j+1} = 2\pi \left(1+\varepsilon\right),
    \label{eq:incomm_spacing}
\end{align}
with the spacing parameter $\varepsilon \ll 1$, the required bond dimension $\chi_\mathrm{max}$ tends to grow faster with the system size, as shown in Fig.~\ref{fig:2}(b). This rapid change in computational cost indicates a significant challenge for the exact TN simulations of arbitrary spin chains with environment-induced long-range interactions. {In the following section we demonstrate another way to solve this problem by using} an alternative quantum many-body state representation based on NNs, \textit{i.e.}, so-called time-dependent neural quantum states (t-NQSs).

\section{Neural quantum states}
\label{sec:NQS}
In this section we present an alternative approach to solve the ME~\eqref{eq:ME} that is based on NNs representing the system state, referred to as NQSs~\cite{carleo_solving_2017}. We base our approach on Ref.~\cite{de_walle_many-body_2024}, where explicitly time-dependent NQSs (t-NQSs) were introduced, and extend this concept to open-system dynamics. Below we provide the reader with the basic ideas behind NQSs as well as important building blocks and computational cost estimates of the t-NQS approach used in this work.

The density matrix $\rho$ can be represented within a NQS framework through various approaches, such as mixed-state purification~\cite{torlai2018latent} or vectorization of the Liouville space~\cite{wei_variational_2025}. In this work, we choose the positive operator-valued measure (POVM) formalism to represent the density matrix~\cite{nielsen_quantum_2010}. In the case of informationally complete POVM (IC POVM), $\rho$ is unambiguously defined via the probability distribution $p\left(\mathbf{a}\right)$:
\begin{align}
    p\left(\mathbf{a}\right) = \mathrm{tr} \left(\rho M_{\left(\mathbf{a}\right)}\right),
    \label{eq:prop_POVM}
\end{align}
where $\mathbf{a}$ spans over $d^{2n}$ configurations and $\left\{M_{\left(\mathbf{a}\right)}\right\}$ is a set of positive semi-definite operators called the IC POVM frame. Since the map~\eqref{eq:prop_POVM} is unambiguous, there exists an inverse transformation that recovers the state, $\rho = \sum_{\mathbf{b}} p\left(\mathbf{b}\right) N^{\left(\mathbf{b}\right)}$, where $\left\{N^{\left(\mathbf{b}\right)}\right\}$ is the dual frame. We refer the reader to Refs.~\cite{carrasquilla_neural_2021, luo_autoregressive_2022} for more information about IC POVM representations in the context of NQSs.

NNs can be viewed as universal multivariate function approximators~\cite{cybenko1989approximation}, and in our case such a function is the distribution $p\left(\mathbf{a}\right)$. The state probabilities from this distribution can be approximated by the NNs in NQSs: $p\left(\mathbf{a}\right) \approx p_\theta\left(\mathbf{a}\right)$, where $\theta$ is a vector of variational NN parameters. When the quantum state $\rho(t)$ evolves in time under the ME~\eqref{eq:ME}, so do the corresponding probabilities:
\begin{align}
    \partial_t {p}\left(\mathbf{a},t\right) = \sum_{\mathbf{b}} A_{\mathbf{a}\mathbf{b}}~p\left(\mathbf{b},t\right),
    \label{eq:EoM_POVM}
\end{align}
where the definition of $A_{\mathbf{a}\mathbf{b}}$ can be found in Appendix~\ref{app:POVM_ME}. There are several ways to encode the time dependence into the NQSs. Pioneering approaches include time‑dependent variational Monte Carlo (t‑VMC)~\cite{carleo2012localization, carleo2017unitary} and projected t‑VMC~\cite{sinibaldi_unbiasing_2023}, which describe the dynamics through time‑dependent NN parameters, $p\left(\mathbf{a},t\right) \approx p_{\theta(t)}\left(\mathbf{a}\right)$. On a practical level this approach implies a separate NN trained for each time step of the evolution. A more recent alternative consists in promoting $t$ to an input parameter of the probability distribution, such that $p\left(\mathbf{a},t\right) \approx p_{\theta}\left(\mathbf{a}, t\right)$. In this case, a single NN is trained globally to represent the quantum state at all times within a given time interval $t\in \left[0,\mathcal T\right]$. This idea was independently introduced by Refs.~\cite{de_walle_many-body_2024, sinibaldi_time-dependent_2024}, with the following key difference. Ref.~\cite{sinibaldi_time-dependent_2024} includes time dependence via a special wavefunction representation called Galerkin ansatz. In turn, Ref.~\cite{de_walle_many-body_2024} incorporates time as an additional input parameter that is being encoded in the NN.

In our work we use the latter method based on autoregressive transformer NNs. {We refer the reader to Appendix~\ref{app:LIC-POVM} for details on autoregressive architectures}. Transformer NNs are universal sequence-to-sequence function approximators initially introduced for language models~\cite{vaswani2017attention}. Their use as NQSs spans a plethora of ground and steady state search methods for various physical models, including impurity models~\cite{cao_vision_2024}, dissipative models~\cite{wei_variational_2025} and frustrated systems like the transverse-field Ising model~\cite{zhang_transformer_2023, lange_transformer_2024}, $J_1$-$J_2$ model~\cite{viteritti2023transformer}, Fermi-Hubbard model~\cite{ibarra2025autoregressive} as well as neutral atom arrays~\cite{sprague2024variational, fitzek2024rydberggpt}. Transformer's (self)-attention mechanism is particularly useful to learn complex long-range correlation patterns in quantum many-body systems, in some cases potentially outperforming existing TN methods~\cite{viteritti2023transformer,lange_transformer_2024, lange_architectures_2024, de_walle_many-body_2024}. The model~\eqref{eq:ME} considered in our work is thus an interesting test case for transformer NQSs, as it involves long-range dissipative interactions that rapidly generate quantum correlations in the system, rendering TN simulations hard, as we have shown in the previous section.

\begin{figure}
\label{fig:3}
    \centering
    \includegraphics[width=0.5\textwidth]{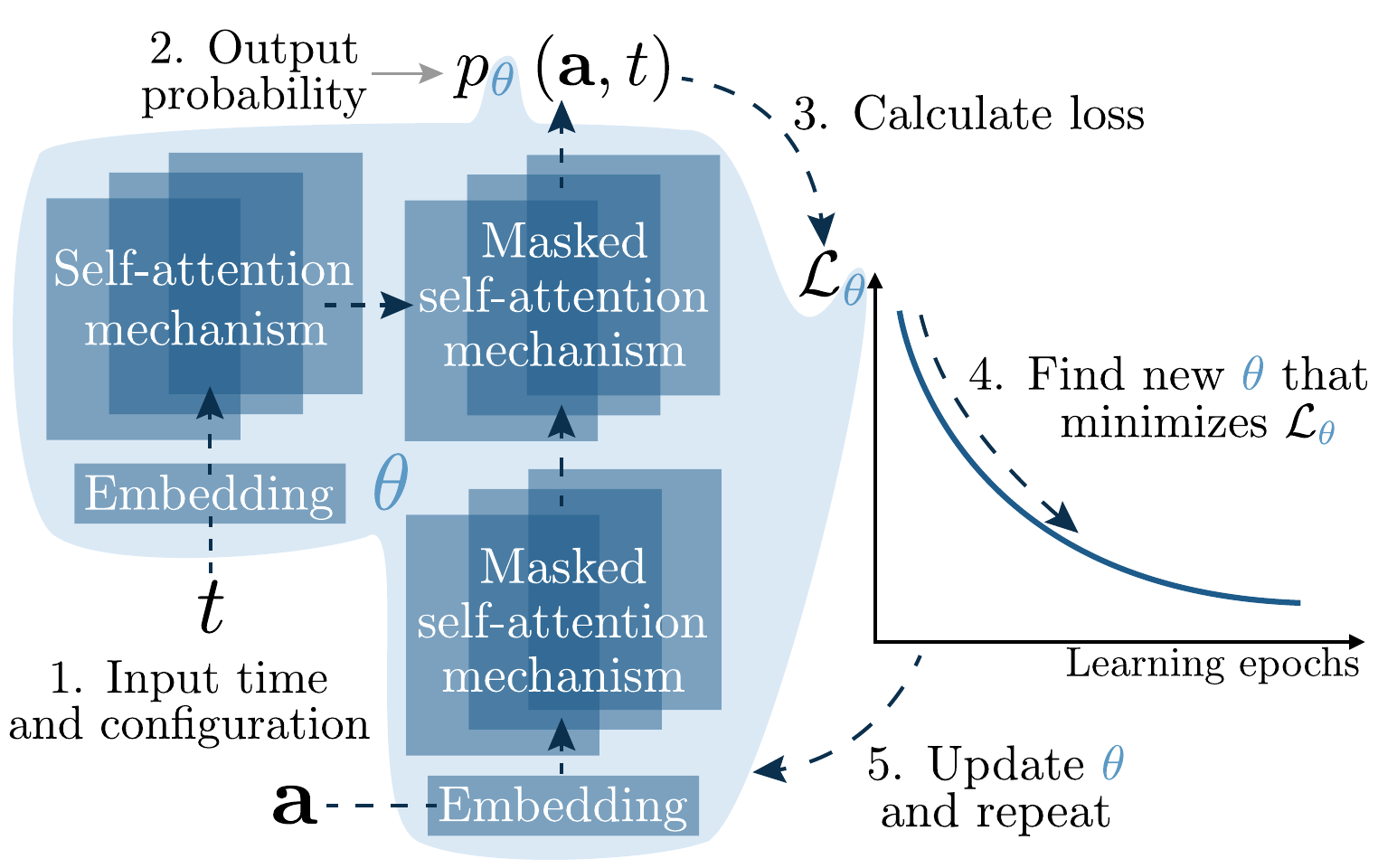}
    \caption{Training routine of the time-dependent NQS (t-NQS) based on an autoregressive transformer NN. The NN with the parameter vector $\theta$ receives the time $t$ and state configuration $\mathbf{a}$ as inputs and produces the corresponding probability distribution approximation $p_\theta\left(\mathbf{a}, t\right)$ as an output. Based on this distribution, the loss function~\eqref{eq:loss} is computed, allowing to perform a single NN optimization step (learning epoch) using standard NN backpropagation techniques. The procedure is repeated until the loss in sufficiently minimized (see Appendix~\ref{app:tech} for further details).
    }
\end{figure}

In Fig.~\ref{fig:3} we show a sketch of the transformer model used to approximate the probability distribution $p\left(\mathbf a, t\right)$. The model has a so-called encoder-decoder structure~\cite{vaswani2017attention}, which allows to naturally separate the configuration input $\mathbf a$ and time input $t$ of the approximate function $p_\theta\left(\mathbf a, t\right)$. The encoder embeds the time input $t$ into a higher-dimensional vector of size $d_h$ called \textit{hidden dimension}. This vector is then passed to an attention layer followed by a feed-forward NN. Within this structure, the model learns correlations in the input and constructs a context vector, which is subsequently passed to the decoder. The decoder takes the input state configuration in the form of $n$ tokens, $\mathbf a = \left(a_1, \dots, a_n\right)$ (see Appendix~\ref{app:LIC-POVM}), and embeds each of them into vectors of size $d_h$. Passing these vectors through attention layers, the decoder integrates them with the time context vector coming from the encoder. This information is then passed through another feed-forward NN with a $\texttt{Softmax}$ activation function and then is finally decoded into an approximation of the probability distribution, $p_\theta\left(\mathbf{a}, t\right)$.

The goal of the NN training is to minimize the \textit{loss function}, which measures the deviation of the NQS output $p_\theta\left(\mathbf a, t\right)$ from the actual distribution $p\left(\mathbf a, t\right)$. In the case of t-NQSs, the loss function is simply the error of Eq.~\eqref{eq:EoM_POVM}. Such a direct loss estimation is possible due to the fact that $t$ is the input of the NN, allowing for an exact and efficient calculation of time derivatives of $p_\theta\left(\mathbf a, t\right)$ using automated differentiation~\cite{10.5555/3454287.3455008}. Once such a loss function is sufficiently globally minimized during the NN training, the resulting t-NQS represents the correct solution of Eq.~\eqref{eq:EoM_POVM}. In the variational Monte Carlo (VMC) framework, this loss function can be written as
\begin{align}
    \mathcal L_\theta = \frac{dt}{|\tau|}\sum_{t\in \tau} \sum_{\mathbf{a}\sim p_\theta\left(\mathbf{a}, t\right)} \left|\frac{\partial_t p_\theta\left(\mathbf{a}, t\right) - \sum_{\mathbf{b}} A_{\mathbf{a}\mathbf{b}}~p_\theta\left(\mathbf{b},t\right)}{p_\theta\left(\mathbf{a}, t\right)}\right|,
    \label{eq:loss}
\end{align}
where the outer sum runs over time steps $t$ from a discrete subset of the whole time interval, $\tau \subset \left[0, \mathcal T\right]$ with $dt$ being the discretization step, and the inner sum involves Born-rule sampling of state configurations. In order to take into account the initial state of the system as well as to simulate the dynamics for several time intervals of duration $\mathcal T$, we utilize the concatenation technique introduced in Ref.~\cite{de_walle_many-body_2024}. We refer the reader to Appendix~\ref{app:tech} for the details on the time interval concatenation, the choice of the interval subset $\tau$ as well as other technical aspects of the t-NQS model.

For the t-NQS model described above, the computational cost is determined by its hidden dimension $d_h$ and the system size $n$~\cite{vaswani2017attention}:
\begin{align}
    \mathcal C_\mathrm{NQS}\left(d_h, n\right) = n^4 d_h + n^3 d_h^2.
    \label{eq:NQS_cost}
\end{align}
The high-order polynomial scaling with $n$ is related to the quadratic scaling of transformer's self-attention and the separate application of two-body terms from the ME~\eqref{eq:ME} during the estimation of the loss~\eqref{eq:loss}. We refer the reader to Appendix~\ref{app:costs} for the detailed definitions of the computational cost. In practice, such a scaling of the cost allows us to perform computations with system sizes up to $n = 20$, as done in Section~\ref{sec:results}.

\section{Results}
\label{sec:results}
\begin{figure}
\label{fig:4}
    \centering
    \includegraphics[width=0.47\textwidth]{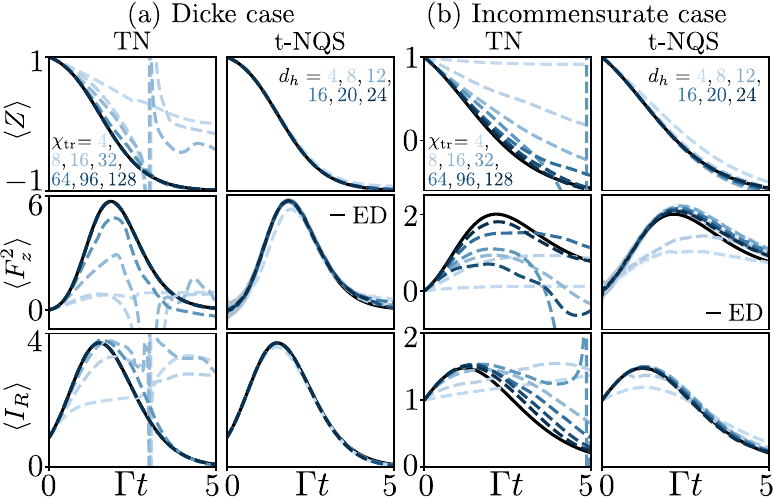}
    \caption{Dynamics of physical observables of a spin chain near a waveguide for (a) commensurate and (b) incommensurate emitter spacings. The first and second rows feature the evolution of magnetization and fluctuations along the $z$-axis, respectively. The third row shows the output intensity in the right-propagating waveguide mode (see main text). Left and right subcolumns in (a) and (b) correspond to TN simulations with various truncation bond dimensions $\chi_\mathrm{tr}$ and t-NQS simulations with varying hidden dimensions $d_h$, respectively. The system size is $n=20$ in (a) and $n=12$ (maximum accessible by the ED in the full Hilbert space) in (b). The shaded regions around t-NQS curves indicate the sampling error.
    }
\end{figure}

\begin{figure}
\label{fig:5}
    \centering
    \includegraphics[width=0.475\textwidth]{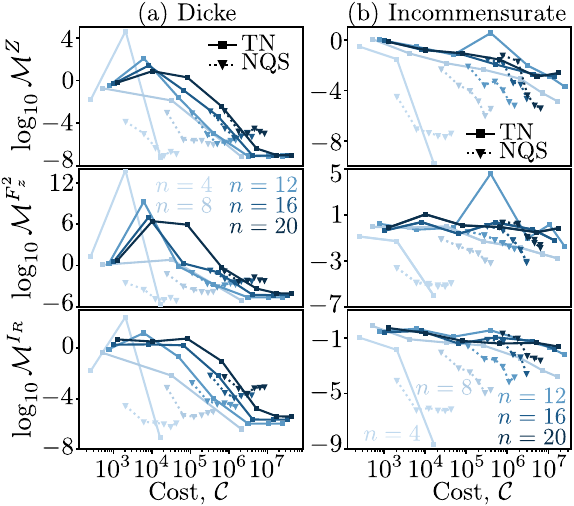}
    \caption{Dependence of the average square deviation on the computational cost values for Dicke (a) and incommensurate (b) scenarios. As in Fig.~\ref{fig:4}, the rows display the magnetization and fluctuations along the $z$-axis (first and second rows), and the right-propagating waveguide mode output (third row). Solid lines with squares and dotted lines with triangles represent average square errors for TN and NQS methods, respectively. The lines' colors indicate different system sizes.
    }
\end{figure}

\begin{figure}
\label{fig:6}
    \centering
    \includegraphics[width=0.45\textwidth]{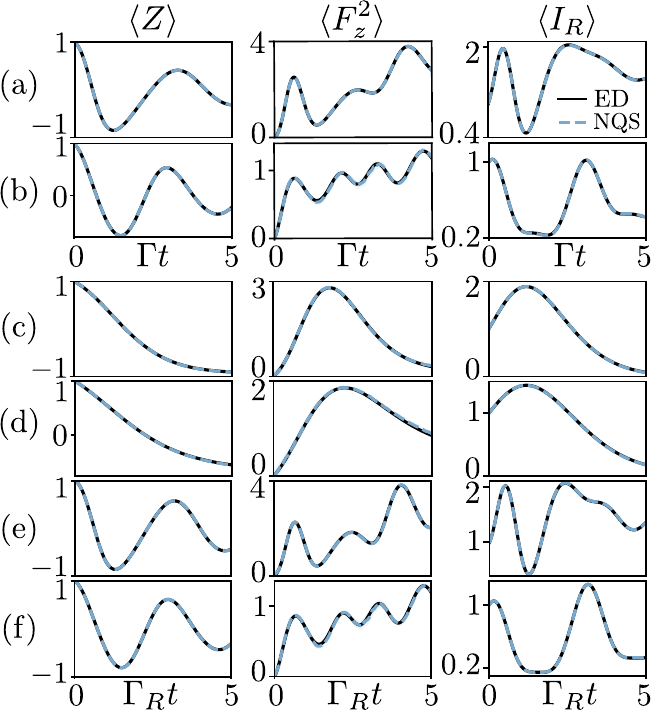}
    \caption{Dynamics of physical observables of the spin chain near a waveguide for several cases of the generalized master equation (ME) as defined in Appendix~\ref{app:ME_gen}: 
    (a) driven Dicke model ($\Gamma_L = \Gamma_R \equiv \Gamma = \Omega$, $k\delta x_{j,j+1}=2\pi$),
    (b) incommensurate case with driving ($\Gamma_L = \Gamma_R \equiv \Gamma = \Omega$, $k\delta x_{j,j+1}=\pi / \sqrt{3}$),
    (c) commensurate chiral case ($\Gamma_L = 0.5 \Gamma_R, \Omega=0$, $k\delta x_{j,j+1}=2\pi$),
    (d) incommensurate chiral case ($\Gamma_L = 0.5 \Gamma_R, \Omega=0$, $k\delta x_{j,j+1}=\pi / \sqrt{3}$),
    (e) commensurate driven chiral case ($\Gamma_L = 0.5 \Gamma_R, \Omega=\Gamma_R$, $k\delta x_{j,j+1}=2\pi$),
    (f) incommensurate driven chiral case ($\Gamma_L = 0.5 \Gamma_R, \Omega=\Gamma_R$, $k\delta x_{j,j+1}=\pi / \sqrt{3}$).
    The columns correspond to the same observables as in Figs.~\ref{fig:4} and~\ref{fig:5}. The system size is $n=8$, the hidden dimension of the t-NQS is $d_h = 16$ for (a), (c), (e) and $d_h = 20$ for (b), (d) and (f). The semi-transparent shadings around the t-NQS plot lines indicate the sampling error.}
\end{figure}

\begin{figure}
\label{fig:7}
    \centering
    \includegraphics[width=0.45\textwidth]{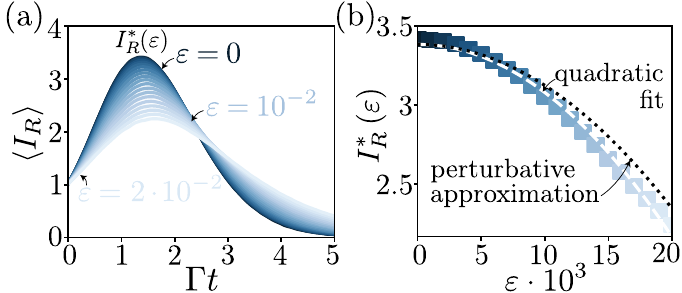}
    \caption{Dependence of the emission intensity dynamics on the emitter spacing. (a) Time evolution of the emission intensity into the right-propagating waveguide mode, $\langle I_R \rangle$, for various emitter spacings (see text). The emission intensity grows during the transient dynamics, reaching the peak value $I_R^*\left(\varepsilon\right)$, and then goes down to zero as the system approaches the steady state. The semi-transparent shadings around the t-NQS plot lines indicate the sampling error. (b) The peak intensity $I_R^*$ as a function of the spacing parameter $\varepsilon$. The perturbative approximation and quadratic fit are shown as dotted black and dashed white lines, respectively. The system size is $n=16$, the hidden dimension of the t-NQS is $d_h = 20$.}
\end{figure}

In this section we compare the performances of the TN method presented in Section~\ref{sec:TN} and the t-NQS method presented in Section~\ref{sec:NQS}. Specifically, we apply both methods to solve the ME~\eqref{eq:ME} that generates the dynamics of a spin-1/2 chain near a 1D waveguide. In Figs.~\ref{fig:4} and~\ref{fig:5} we compare the convergence properties of both methods for two instances of the balanced ME: the commensurate (Dicke) regime with $k \delta x_{j,j+1} = 2\pi$ and the incommensurate regime with $k \delta x_{j,j+1}  = \pi/\sqrt{3}$. In addition, in Fig.~\ref{fig:6} we demonstrate the capability of the t-NQS method to solve various instances of the general ME (as defined in Appendix~\ref{app:ME_gen}). Finally, using the t-NQS simulations, in Fig.~\ref{fig:7} we consider how the waveguide QED physics changes once moving away from the permutationally symmetric Dicke regime.

We consider how several system observables evolve under the ME~\eqref{eq:ME}. Specifically, we consider expectation values of the magnetization and fluctuations (cumulative two-body correlations) along the $z$ direction of the spin chain:
\begin{align}
    &\langle Z \rangle \equiv \frac{1}{n}\sum_{j=1}^n \langle \sigma_j^z \rangle,\\
    &\langle F^2_z \rangle \equiv \frac{1}{n}\sum_{j,\ell=1}^n \left(\langle \sigma_j^z \sigma_\ell^z \rangle - \langle \sigma_j^z \rangle \langle \sigma_\ell^z \rangle\right).
\end{align}
as well as the intensity $\langle I_R \rangle \equiv \langle c_{R}^\dagger c_{R} \rangle / n $ of the output emission in the right direction of the waveguide governed by the jump operator $c_{R} = \sum_{j} e^{ik x_j} \sigma^-_j$ (see Appendix~\ref{app:ME_gen} and Ref.~\cite{pichler_quantum_2015} for details). This choice allows us to probe the performances of our numerical methods for both linear (magnetization) and quadratic (fluctuations, output intensity) observables.

In the Dicke case shown in Fig.~\ref{fig:4}(a) we observe a stable convergence by both the TN and t-NQS methods for all of the chosen observables. Specifically, the TN method coincides with the ED at all times already for the truncation\footnote{Truncation bond dimension $\chi_\mathrm{tr}$ is the maximal bond dimension allowed during simulations.} bond dimension $\chi_\mathrm{tr} = 64$, which confirms the polynomial scaling from Fig.~\ref{fig:2}(b). Incidentally, the t-NQS method gives correct predictions at a rather low hidden dimension $d_h = 8$. Moreover, we observe a good approximation with even lower values of $\chi_\mathrm{tr}$ or $d_h$ for the magnetization $\langle Z \rangle$.

The incommensurate case is different from the Dicke regime, as can be seen from comparing Fig.~\ref{fig:4}(b) with Fig.~\ref{fig:4}(a). Specifically, for both methods the convergence to the ED is slower for all considered observables. In the context of TNs, this is consistent with the finding shown in Fig.~\ref{fig:2}(b) of Section~\ref{sec:TN}. In this case, one can also observe a slightly better convergence for the linear observable in comparison to the quadratic observables for both methods. On the other hand, the t-NQS lines approach the ED lines closer, even if not converging as good as in the Dicke case. It is important to note that increasing the truncation bond dimension $\chi_\text{tr}$ of TNs and the hidden dimension $d_h$ of NQSs leads to different scaling of the associated computational costs, as indicated in Eqs.~\eqref{eq:MPO_cost} and~\eqref{eq:NQS_cost}. For instance, increasing $\chi_\text{tr}$ from 4 to 8 results in an eightfold increase in computational cost $\mathcal C_\text{TN}$, whereas increasing $d_h$ from 4 to 8 leads to only a fourfold increase of $\mathcal C_\text{NQS}$. In light of this, together with the results of Fig.~\ref{fig:4}, we find that for a comparable increase in computational cost, the NQSs exhibit faster convergence than the TNs in the specific scenarios considered.

The results of Fig.~\ref{fig:4} motivate comparing the convergence properties of the studied numerical methods. As a measure of how well the simulation converges for a given observable $\langle O \rangle$, we take the time average of the squared deviation of the numerical data from the ED solution for a fixed evolution time $T$:
\begin{align}
    \mathcal M^O_\mathrm{num} \left(\mathcal C \right) = \frac{1}{T} \int_0^{T} dt~\left(\langle O \rangle_\mathrm{num}(t) - \langle O \rangle_\mathrm{ED} (t)\right)^2,
    \label{eq:conv}
\end{align}
where the subscript $\mathrm{num} \in \left\{\mathrm{TN},~\mathrm{NQS}\right\}$ indicates the numerical method, $O$ is the operator corresponding to a physical observable and $\mathcal C$ is the corresponding computational cost as defined in Eqs.~\eqref{eq:MPO_cost} and~\eqref{eq:NQS_cost}. We note that the error~\eqref{eq:conv} stems from two different sources. The first is the error of the state approximation that is defined in TNs by the truncation bond dimension $\chi_\mathrm{tr}$ and in NQSs by the hidden dimension $d_h$. The second source is related to all residual errors, \textit{e.g.}, an error from a finite time step in the first-order time propagation in TNs or finite sampling size, limited optimization capabilities and time interval discretization in t-NQSs. In the following analysis we focus on the former error of the state approximation, while keeping the residual errors small. Residual errors thus set the lower limit for the value of the average square deviation~\eqref{eq:conv}. This means that, upon reducing the state approximation error by increasing the associated computational cost $\mathcal C$, the value of $\mathcal M^O_\mathrm{num}$ reduces and eventually saturates at the value of the residual error. 

We therefore aim to find out how much computational cost does a given state ansatz require to reach the threshold set by the residual errors. We thus state that the simulation is converged in terms of the state approximation for a given observable $O$ if the average square error $\mathcal M^O_\mathrm{num}$ reaches a plateau at some sufficient cost $\mathcal C^*$,
\begin{align}
    \mathcal M^O_\mathrm{num} \left(\mathcal C \geq \mathcal C^*\right) = \tilde{\mathcal M}^O_\mathrm{num},
    \label{eq:conv_cond}
\end{align}
where the value of the plateau $\tilde{\mathcal M}^O_\mathrm{num}$ is set by the residual errors.

In Fig.~\ref{fig:5} we analyze the reduction the overall average square deviations of both methods, $\mathcal M^O_\mathrm{TN}$ and $\mathcal M^O_\mathrm{NQS}$, in terms of increasing computational cost\footnote{We note that the costs for TNs and NQSs from Eqs.~\eqref{eq:MPO_cost} and~\eqref{eq:NQS_cost} are defined as scalings in different relative units and should not be compared quantitatively. Moreover, the considered ranges of computational costs differ for TNs and NQSs. This stems from our choice of the bond dimensions $\chi_\mathrm{tr}$ and hidden dimensions $d_h$.}. The TN and NQS average square deviations are indicated by solid and dashed lines, respectively. We consider the Dicke and incommensurate regimes in Fig.~\ref{fig:5}(a) and (b) and indicate various system sizes $n$ by different colors. We note that in the Dicke case, the average square error is evaluated against the ED for all system sizes. In the incommensurate case, the ED is used for the system sizes up to $n=12$, while for larger $n$, $\mathcal M^O_\mathrm{TN}$ and $\mathcal M^O_\mathrm{NQS}$ are calculated relative to the solutions with the largest bond dimension $\chi_\mathrm{tr}$ or hidden dimension $d_h$, respectively.

From the Dicke case shown in Fig.~\ref{fig:5}(a), one can conclude that both methods satisfy the convergence conditions. One can observe that the TN average square error is smaller compared to the NQS case, $\tilde{\mathcal M}^O_\mathrm{TN} < \tilde{\mathcal M}^O_\mathrm{NQS}$. This is expected because in this case the TNs represent the state exactly up to the second order in the time step~\cite{zaletel_time-evolving_2015}, while the residual errors in NQSs are larger and depend on the system size\footnote{The system-size dependence of the residual errors in NQSs can be explained by the quadratically growing number of Lindbladian terms in the loss~\eqref{eq:loss} and henceforth increased learning effort with the system size, and also by the discretization of time intervals in t-NQSs.}. A completely different scenario can be seen for the incommensurate regime shown in Fig.~\ref{fig:5}(b). Here, only for the smallest system size, $n=4$, the TNs provide a solution that is close to the ED\footnote{{For TNs with low bond dimensions one can observe a larger deviation for the Dicke model compared to the incommensurate case. This can potentially be related to the {non-positivity} of the TN solution upon truncation of the bond dimension~\cite{verstraete2004matrix, kliesch2014matrix}.}}, while for larger system sizes the TN solutions fail to satisfy the convergence condition~\eqref{eq:conv_cond} for all of the observables. In turn, the t-NQS solutions formally satisfy~\eqref{eq:conv_cond} for the cases of the magnetization $\langle Z \rangle$ and output intensity $\langle I_R \rangle$, and in the case of fluctuations $\langle F^2_z \rangle$ t-NQSs confidently exhibit lower deviation values with increasing cost. Even though the deviation here is larger compared to the one in the Dicke case, it can be confirmed that for all observables, the deviations accessible by NQSs are \textit{lower} than the deviations accessible by TNs, $ \mathcal M^O_\mathrm{NQS} (\mathcal C_\mathrm{NQS}^* ) < \mathcal M^O_\mathrm{TN} (\mathcal C_\mathrm{TN}^* )$, with $\mathcal C_\mathrm{num}^*$ being the maximal cost accessible by the respective numerical method. This confirms that, compared to the simulations accessible by the TN method, the t-NQS predictions are more accurate for incommensurate spin dynamics, while demonstrating no substantial gain in the case of permutationally symmetric Dicke model.

With the t-NQS method presented here, one is able to study the dynamics of various waveguide QED settings beyond the examples considered so far. In Fig.~\ref{fig:6} we take several instances of the general ME (presented in Appendix~\ref{app:ME_gen}) and plot the dynamics of different physical observables generated by the t-NQS method, comparing them to the ED results. One can see that the t-NQS is capable of capturing complex dynamics of driven systems with collective decay as well as emitters coupled to a chiral waveguide. We note that the incommensurate cases shown in Fig.~\ref{fig:6}(b), (d), and (f), while requiring a higher hidden dimension ($d_h = 20$) compared to the commensurate case ($d_h = 16$), converge smoothly to the ED curves, demonstrating the ability of the t-NQS to accurately represent the dynamics even in more complex settings.

Using the t-NQS method, one can explore the physics of the waveguide QED beyond the standard Dicke regime~\cite{holzinger_beyond_2025}. Specifically, in Fig.~\ref{fig:7} we consider the dynamics of the emission intensity $\langle I_R \rangle$ and how the peak intensity value $I_R^*$ scales with the emitter spacing once deviating from the commensurate regime. This deviation is parametrized by $\varepsilon$ in Eq.~\eqref{eq:incomm_spacing} and is also studied with TNs in Fig.~\ref{fig:2}(b). Fig.~\ref{fig:7}(a) shows that the parameter $\varepsilon$ has a pronounced impact on the dynamics of $\langle I_R \rangle$, with growing $\varepsilon$ leading to a gradual suppression of the intensity peak $I_R^*$. Within the perturbation-theoretic approximation\footnote{We refer the reader to Appendix~\ref{app:pert} for the derivation of the perturbative approximation of the quadratic scaling.}, one can expect $I_R^*$ to decrease quadratically with sufficiently small $\varepsilon$, which is indicated by a black dotted line in Fig.~\ref{fig:7}(b). We confirm this behavior with our t-NQS simulations, the results of which appear as square data points in Fig.~\ref{fig:7}(b). We additionally provide a quadratic fit\footnote{We refer the reader to Appendix~\ref{app:pert} for the parameters of the numerical quadratic fit.} of the data, shown as a white dashed line in Fig.~\ref{fig:7}(b).
The results presented in Fig.~\ref{fig:7} indicate that a slight deviation from commensurate emitter spacings, a common occurrence in experimental settings, leads to a quadratic suppression of the superradiant emission rate.

\section{Conclusions and Outlook}
\label{sec:conclusions}
In this work, we introduce a novel neural quantum state (NQS) approach for simulating the dynamics of spin chains with dissipative all-to-all interactions mediated by a one-dimensional Markovian waveguide. Our method builds on the recently proposed time-dependent NQS (t-NQS) framework, which incorporates time as an explicit input to the neural network~\cite{de_walle_many-body_2024}, extending the approach to the density matrix representation using informationally complete positive operator-valued measures (IC POVM)~\cite{carrasquilla_neural_2021}. We benchmark the performance of t-NQS against a conventional tensor network (TN) method based on the matrix product state (MPS) representation of the vectorized density matrix~\cite{schollwock_density_2011, wellnitz2022rise, preisser_comparing_2023}. {Our results demonstrate that the t-NQS method is able to capture spin dynamics in generic, symmetry-breaking configurations, albeit currently limited to system sizes of up to $n= 20$. }{In particular, our method predicts the dynamics of the incommensurate emitter spacings more accurately than the TN approach across accessible bond dimensions, while performing on par with TNs in the permutationally symmetric Dicke regime.} Additionally, we demonstrate that the developed numerical method is able to simulate various physical regimes of the waveguide QED, including emitter systems with driving, incommensurate spacings and chiral dissipative couplings. Finally, with the developed tool we study how the phenomenon of superradiance changes once moving away from the permutationally symmetric regime, and observe that the peak emission intensity is quadratically suppressed with increasing perturbation in the emitter spacings, in agreement with perturbative predictions.

Several directions remain open for future research. From the physics perspective, an important next step is to further investigate the superradiance in symmetry-broken regimes. In particular, the perturbative deviation from the Dicke model presented in Fig.~\ref{fig:7} could be extended using finite-size scaling argument, to characterize the fate of superradiance in the thermodynamic limit. On the methodological side, the t-NQS framework developed in this work can naturally accommodate a broader class of physical models. For instance, our implementation readily extends to systems with time-dependent parameters, enabling the study of complex processes such as optimal control and dissipative state preparation~\cite{stannigel2012driven, pocklington2022stabilizing, pocklington_accelerating_2024}. Further improvements may come from exploring alternative NQS representations. Specifically, employing neural network ans\"atze operating on the vectorized density matrix~\cite{wei_variational_2025} could offer advantages over the POVM-based sampling used in this work. Moreover, the vectorized form of the NQS could open doors for developing mappings between the TN and NN representations of quantum states~\cite{wu_tensor-network_2023}, facilitating switching between different methods in order to overcome entanglement barriers such as ones shown in Fig.~\ref{fig:2}(a). A key technical challenge in this direction remains the efficient treatment of all-to-all interactions within the NQS framework. Addressing this will be essential for scaling to larger system sizes and more complex geometries. Finally, while we compared our approach with the standard TN representation of the vectorized density matrix~\cite{schollwock_density_2011, wellnitz2022rise, preisser_comparing_2023}, we did not consider quantum trajectory based TN approaches~\cite{daley_quantum_2014, manzoni_simulating_2017, vovk_entanglement-optimal_2022, cabot_quantum_2023, vovk2024quantum}. A systematic comparison with such methods remains an important future research direction. Taken together, these directions suggest that NQS-based methods may offer a promising route for simulating open quantum many-body dynamics, especially as their flexibility and scalability are further explored.

\section*{Acknowledgments}
We thank Ana Maria Rey, Darrick Chang, Ana Asenjo-Garcia, Alexander Poddubny, Ephraim Shahmoon, Markus Schmitt, Dominik Wild, Luca Arceci, Fan Yang, Giuliano Giudici, Katherine Van Kirk, Bennet Windt and Nikita Leppenen for fruitful discussions. The computational results presented here have been achieved using the LinuX Cluster of the Institute for Theoretical Physics of the University of Innsbruck, LEO HPC infrastructure of the University of Innsbruck and the {HPC infrastructure of the Arnold Sommerfeld Center for Theoretical Physics at the Ludwig Maximilian University in Munich}. This work is supported by the Deutsche Forschungsgemeinschaft (DFG, German Research Foundation) under Germany’s Excellence Strategy (EXC-2111 -- 390814868), the European Union’s Horizon Europe research and innovation program under Grant Agreement No. 101113690 (PASQuanS2.1), the ERC Starting grant QARA (Grant No. 101041435), the EU-QUANTERA project TNiSQ (N-6001), the Austrian Science Fund (FWF) (Grant No. DOI 10.55776/COE1). For open access purposes, the author has applied a CC BY public copyright license to any author accepted manuscript version arising from this submission.

\bibliography{bibliography} 

\clearpage
\pagebreak
\widetext
\begin{center}
\textbf{\large APPENDICES}
\end{center}
\setcounter{equation}{0}
\setcounter{figure}{0}
\setcounter{table}{0}
\setcounter{section}{0}
\setcounter{footnote}{0}
\makeatletter
\renewcommand{\theequation}{S\arabic{equation}}
\renewcommand{\thefigure}{S\arabic{figure}}
\renewcommand{\bibnumfmt}[1]{[S#1]}
\section{General master equation for the imbalanced Markovian waveguide}
\label{app:ME_gen}
In the case when the waveguide has different dissipation rates for the left- and right-propagating modes, $\Gamma_L \neq \Gamma_R$, the dissipation-induced terms of the ME~\eqref{eq:ME} can be written as sums over the left and right contributions:
\begin{align}
    &H_\mathrm{diss} = H_L+H_R, \label{eq:H_diss_imbalanced} \\
    &\mathcal D\left[\rho\right] = D_L\left[\rho\right] + D_R\left[\rho\right], \label{eq:D_imbalanced}
\end{align}
where the coherent contributions~\eqref{eq:H_diss_imbalanced} have the following form:
\begin{subequations}
    \begin{align}
        &H_L = -\frac{i \gamma_L}{2} \sum_{j<\ell} \left(e^{ik\left|x_j - x_\ell\right|} \sigma^+_j \sigma^-_\ell-\mathrm{H.c.}\right), \label{eq:H_L} \\
        &H_R = -\frac{i \gamma_R}{2} \sum_{j>\ell} \left(e^{ik\left|x_j - x_\ell\right|} \sigma^+_j \sigma^-_\ell-\mathrm{H.c.}\right), \label{eq:H_R}
    \end{align}
\end{subequations}
with $\gamma_{L,R} = \Gamma_{L,R} / n$ are the decoherence rates rescaled with $n$~\cite{carollo_exact_2022, cabot_quantum_2023}. The dissipators~\eqref{eq:D_imbalanced} read:
\begin{align}
    D_{\alpha}\left[\rho\right] = \gamma_{\alpha} \left(c_{\alpha} \rho c_{\alpha}^\dagger - \frac{1}{2}\left\{c_{\alpha}^\dagger c_{\alpha}, \rho\right\}\right),
\end{align}
where $\alpha\in\left\{L,R\right\}$ and $c_{L,R} = \sum_{j} e^{\mp ik x_j} \sigma^-_j$. We also refer the reader to Refs.~\cite{pichler_quantum_2015, suarez-forero_chiral_2024} for further information on chiral spin networks.

\section{Tensor-network master equation propagator}
\label{app:MPO_prop}
A site-$j$ tensor from an MPO that encodes the propagation under the balanced ME~\eqref{eq:ME}--\eqref{eq:H_D_balanced} can be written as~\cite{crosswhite2008finite, hubig2017generic}
\begin{widetext}
\begin{align}
\label{eq:MPO_prop}
    A_j =
    \begin{pmatrix}
        \mathbb I & c^{\rightarrow}_{-} & c^{\rightarrow}_{+} & (c^{\rightarrow}_{-})^\dagger & (c^{\rightarrow}_{+})^\dagger & 
        c^{\leftarrow}_{-} & c^{\leftarrow}_{+} & (c^{\leftarrow}_{-})^\dagger & (c^{\leftarrow}_{+})^\dagger & \mathcal L_\mathrm{loc}\\
        0 & \mathbb I & 0 & 0 & 0 & 0 & 0 & 0 & 0 & - (c^{\rightarrow}_{-})^\dagger + (c^{\leftarrow}_{-})^\dagger \\
        0 & 0 & \mathbb I & 0 & 0 & 0 & 0 & 0 & 0 & (c^{\leftarrow}_{+})^\dagger \\
        0 & 0 & 0 & \mathbb I & 0 & 0 & 0 & 0 & 0 & c^{\rightarrow}_{-} / 2 - c^{\rightarrow}_{-} / 2 \\
        0 & 0 & 0 & 0 & \mathbb I & 0 & 0 & 0 & 0 & - c^{\rightarrow}_{+} \\
        0 & 0 & 0 & 0 & 0 & \mathbb I & 0 & 0 & 0 & (c^{\leftarrow}_{-})^\dagger / 2 - (c^{\leftarrow}_{-})^\dagger / 2 \\
        0 & 0 & 0 & 0 & 0 & 0 & \mathbb I & 0 & 0 & - (c^{\leftarrow}_{+})^\dagger \\
        0 & 0 & 0 & 0 & 0 & 0 & 0 & \mathbb I & 0 & (- c^{\leftarrow}_{-} + c^{\rightarrow}_{-}) \\
        0 & 0 & 0 & 0 & 0 & 0 & 0 & 0 & \mathbb I & c^{\rightarrow}_{+} \\
        0 & 0 & 0 & 0 & 0 & 0 & 0 & 0 & 0 & \mathbb I \\
    \end{pmatrix}_j,
\end{align}
\end{widetext}
where $\mathbb I = \mathbb{1} \otimes \mathbb{1}$ is the identity operator in the Liouville space, $c^{\rightarrow}_{\pm} = \sqrt{\gamma dt} \left(\sigma^- \otimes \mathbb{1}\right) e^{\pm i k x}$ and $c^{\leftarrow}_{\pm} = \sqrt{\gamma dt} \left(\mathbb{1} \otimes \sigma^- \right) e^{\pm i k x}$ are the short notations for jump operators in the Liouville space, and the local Lindbladian term being
\begin{align}
    \mathcal L_\mathrm{loc} &= -i \Omega dt \left(\sigma^x \otimes \mathbb{1} - \mathbb{1} \otimes \sigma^x\right) \nonumber\\
    &+ 2\gamma dt \left(\sigma^- \otimes \sigma^+ - \sigma^+ \sigma^- \otimes \mathbb{1} - \mathbb{1} \otimes \sigma^+ \sigma^- \right) \nonumber.
\end{align}
Note that the site index $j$ is omitted but implied for every operator appearing in~\eqref{eq:MPO_prop}. The resulting MPO has a fixed bond dimension $\chi_\mathrm{prop} = 10$ for any system size and is able to encode both permutationally symmetric (Dicke) and incommensurate propagators. In the former case, the exponents in jump operators $c^{\rightarrow}_{\pm}$ and $c^{\leftarrow}_{\pm}$ correspond to a trivial gauge and can be safely dropped, thus leading to identical MPO tensors that do not depend on the position $j$. The generalization to the \textit{imbalanced} ME presented in Appendix~\ref{app:ME_gen} is straightforward.

\section{Master equation as an equation of motion for probability distributions}
\label{app:POVM_ME}
The balanced ME~\eqref{eq:ME}--\eqref{eq:H_D_balanced} for the density matrix $\rho(t)$ can be rewritten in IC POVM representation as Eq.~\eqref{eq:EoM_POVM}, which is an equation of motion for the probability distribution $p\left(\mathbf{a},t\right)$ defined in Eq.~\eqref{eq:prop_POVM}. In this ME, the generator of time evolution consists of coherent and incoherent contributions, $A_{\mathbf{a}\mathbf{b}} = C_{\mathbf{a}\mathbf{b}} + D_{\mathbf{a}\mathbf{b}}$, which can be written as follows:
\begin{align}
    &C_{\mathbf{a}\mathbf{b}} = -i ~\mathrm{tr}\left(H+H_\mathrm{diss}, \left[N^{\left(\mathbf{b}\right)}, M_{\left(\mathbf{a}\right)}\right]\right),\\
    &D_{\mathbf{a}\mathbf{b}} = {{\gamma}} \sum_{j,\ell} \cos\nu_{j\ell}~\mathbb{D}^{\mathbf{ab}}_{j\ell},\nonumber\\
    &\mathbb{D}^{\mathbf{ab}}_{j\ell} = \mathrm{tr} \left(2\sigma_\ell^- N^{\left(\mathbf{b}\right)} \sigma_j^+ M_{\left(\mathbf{a}\right)} - \sigma_j^+ \sigma_\ell^- \left\{N^{\left(\mathbf{b}\right)}, M_{\left(\mathbf{a}\right)}\right\}\right),\nonumber
\end{align}
where $\mathbf{a}$ and $\mathbf{b}$ are the state configuration vectors, $M_{\left(\mathbf{a}\right)}$ and $N^{\left(\mathbf{b}\right)}$ are the IC POVM frame and dual frame, respectively, and definitions of the remaining terms can be found in the discussion to Eqs.~\eqref{eq:ME} and~\eqref{eq:H_D_balanced} of the main text. The generalization to the \textit{imbalanced} ME presented in Appendix~\ref{app:ME_gen} is straightforward.

\section{Local probability representations}
\label{app:LIC-POVM}
As mentioned in the main text, the t-NQSs used in this work are based on autoregressive NN sampling. This approach consists in representing the NN output, \textit{i.e.}, the probability distribution $p\left(\mathbf{a}\right)$, in a local basis:
\begin{align}
p\left(\mathbf{a}\right) = p\left(a_1, \dots, a_n\right),
\end{align}
where each $a_i$ spans $d^2 = 4$. Such a representation is generated with local IC POVM frames, $\left\{M_{\left(\mathbf{a}\right)}\right\} = \left\{M_{\left(a_1\right)}\otimes \dots \otimes M_{\left(a_n\right)}\right\}$. Each of the local frames $M_{\left(a_j\right)}$ can be chosen in multiple ways as long as it corresponds to an IC POVM representation. In this work we choose each of the local frames to form a so-called Pauli-4 basis~\cite{luo_autoregressive_2022}, which can be written as follows:
\begin{align}
    &M_{(0)_j} = \frac{1}{3}\ket{0}_j\bra{0},\\
    &M_{(1)_j} = \frac{1}{3}\ket{+}_j\bra{+},\nonumber\\
    &M_{(2)_j} = \frac{1}{3}\ket{r}_j\bra{r},\nonumber\\
    &M_{(3)_j} = \mathbb{1}_j - M_{(0)_j} - M_{(1)_j} - M_{(2)_j}.\nonumber
\end{align}
Here $\ket{0}_j$, $\ket{+}_j$ and $\ket{r}_j$ are the eigenvectors with eigenvalue $+1$ of spin operators $\sigma_j^x$, $\sigma_j^y$ and $\sigma_j^z$, respectively ($j^\mathrm{th}$ spin Pauli operators in the $x$, $y$ and $z$ direction).

This allows to subsequently sample local configurations from conditional probabilities, $p\left(\mathbf{a}\right) = \prod_{j=1}^n p\left(a_j | a_1,\dots,a_{j-1}\right)$, and avoid global sampling with computationally costly Markov chain Monte Carlo techniques~\cite{vicentini_variational_2019, luo_autoregressive_2022}. Such an economic sampling is possible in our case, since the probability distributions remain normalized under the trace-preserving ME~\eqref{eq:ME}. The NN architectures suitable for autoregressive sampling include recurrent NNs (RNNs)~\cite{carrasquilla_reconstructing_2019, donatella_dynamics_2023, nys_real-time_2024}, masked convolutional neural networks (CNNs)~\cite{van2016conditional, carrasquilla_neural_2021, wu_tensor-network_2023} and transformers~\cite{carrasquilla_probabilistic_2021, luo_autoregressive_2022, zhang_transformer_2023, sprague2024variational, lange_transformer_2024}, the latter being used in the present work.

\section{Technical details on the neural network model}
\label{app:tech}
The t-NQS described in Section~\ref{sec:NQS} is trained to represent the mixed state of the system at all times $t$ in a given time interval $t\in\left[0,\mathcal T\right]$. Since the intermediate dynamics depends on the initial state $\rho_0$, it is crucial to set up the model at the correct state at time $t = 0$. We follow Ref.~\cite{de_walle_many-body_2024} by incorporating an initialization (concatenation) layer at the end of the NN architecture, which blends fixed initial IC POVM probabilities $p_0\left(\mathbf{a}\right) = \mathrm{tr}\left(\rho_0 M_{\left(\mathbf{a}\right)}\right)$ with the output values of the trained NN with parameters $\theta_1$. This layer forces the total t-NQS output to have the form of a simple convex combination in time domain:
\begin{align}
    &\mathrm{concat}\left(p_0, p_{\theta_1},t\right)=f\left(t\right) p_0\left(\mathbf{a}\right) + \left(1 - f\left(t\right)\right) p_{\theta_1}\left(\mathbf{a}, t\right)\nonumber \\
    &\text{with~}t\in\left[0,\mathcal T\right],
    \label{eq:concat}
\end{align}
where $f\left(t\right)$ is an arbitrary differentiable function with boundary conditions $f\left(0\right) = 1$ and $f\left(\mathcal T\right)=0$, and $p_{\theta_1}\left(\mathbf{a}, t\right)$ corresponds to the output of the NN parametrized by $\theta_1$. For all the results presented in this work, we choose the simplest concatenation function, $f\left(t\right) = 1 - t$, as we found it to facilitate the NN training. 

We note that this approach can be extended beyond a single time interval, \textit{e.g.}, it can be used to concatenate two subsequent intervals. In this case, the concatenation layer blends the previous NN output, $p_{\theta_1}\left(\mathbf{a}, \mathcal T\right)$, taken at the last time step of its interval, and the next network, $p_{\theta_2}\left(\mathbf{a}, t\right)$, at the subsequent interval $t\in\left[\mathcal T, 2\mathcal T\right]$:
\begin{align}
&\mathrm{concat}\left(p_{\theta_1},p_{\theta_2},t\right)=f\left(t\right) p_{\theta_1}\left(\mathbf{a}, \mathcal T\right) + \left(1 - f\left(t\right)\right) p_{\theta_2}\left(\mathbf{a}, t\right)\nonumber \\
    &\text{with~}t\in\left[\mathcal T,2\mathcal T\right].
    \label{eq:concat2}
\end{align}
The first NN is treated as an ``initial state'' and its parameters $\theta_1$ are fixed, \textit{i.e.}, there is no training occurring for this NN, and the second NN with parameters $\theta_2$ is being trained for the corresponding time interval. In this way one can facilitate the training for a larger time range, $\left[0, m\mathcal T\right]$ with $m\in\mathbb{Z}^{+}$, by dividing the range into $m$ smaller intervals of duration $\mathcal T$, training a separate NN for each of them and concatenating them one after another in a sequence.

\begin{figure}
\label{fig:SM_1}
    \centering
    \includegraphics[width=0.4\textwidth]{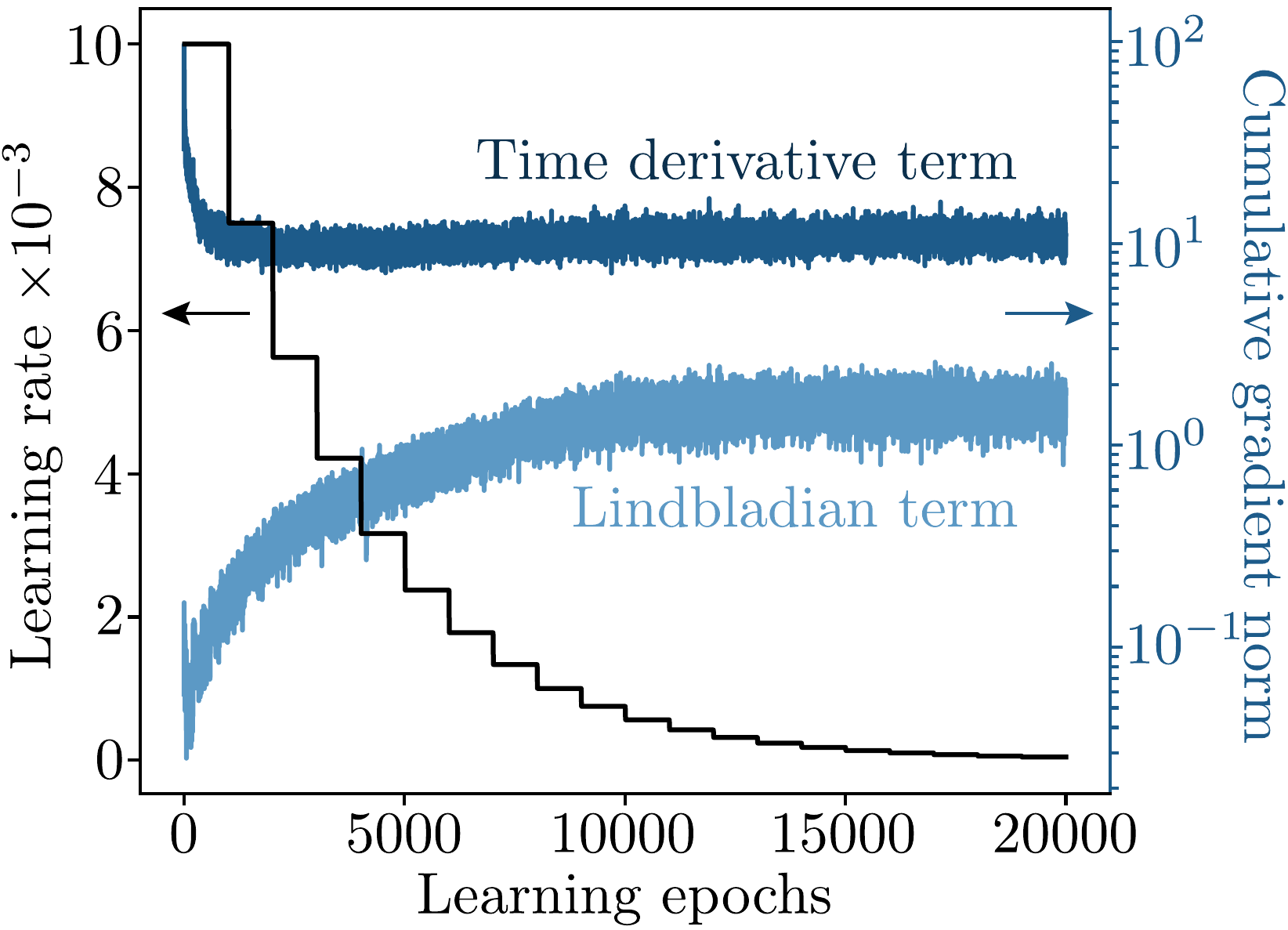}
    \caption{A typical example of the gradient absolute value accumulated for all the NN parameters $\theta$ for the time derivative (dark blue solid line) and Lindbladian (light blue solid line) contributions to the loss~\eqref{eq:loss} during the training of a single t-NQS. The black solid line indicates the learning rate changes during the training. Here the physical parameters are: the system size is $n=4$, the distance between the emitters is $k \left|x_j - x_{j+1}\right| = \pi/\sqrt{3}$, the driving is $\Omega=0.2 \Gamma_R$, and decoherence rates are $\Gamma_L=0.5 \Gamma_R$. The NN model has hidden dimension $d_h = 20$, with the rest of the model parameters being the same as in the text.
    }
\end{figure}

The t-NQS output at a given interval is used to estimate the loss function $\mathcal L_\theta$ provided in Eq.~\eqref{eq:loss} of the main text. The training of the corresponding transformer NN consists in minimization of $\mathcal L_\theta$ with respect to the NN parameters $\theta$. This minimization is performed using the gradient-based backpropagation in the PyTorch framework~\cite{10.5555/3454287.3455008}. We find that, for the purpose of optimization, it is sufficient to include only the gradient of the first term in the loss function~\eqref{eq:loss} that corresponds to the time derivative, while neglecting the gradient contribution from the second, Lindbladian term. The reason for this is evident from Fig.~\ref{fig:SM_1}, where we plot cumulative absolute values of gradients summed across all the NN parameters $\theta$ of the corresponding loss terms. These values are simple indicators of the importance of the corresponding gradient contributions for the overall learning process. Additionally we plot the learning rate schedule (see further technical details in the next paragraph). We observe a clear dominance of the time derivative term in the beginning of training when the learning rate is the highest. Upon the end of training, when the model is no longer learning efficiently, the gradients of the time derivative and Lindbladian contributions are still separated by an order of magnitude. This justifies the described gradient approximation, and we find that our method still successfully minimizes the loss~\eqref{eq:loss} and converges to the correct dynamics. This heuristic significantly reduces memory consumption, thereby enabling the simulation of larger system sizes.

For minimization of the loss~\eqref{eq:loss} we use the standard ADAM optimizer for the NN optimization with multi-step learning rate scheduler~\cite{10.5555/3454287.3455008}. The scheduler parameters are as follows: total number of epochs is $N_\mathrm{epochs} = 2\times10^4$, the learning milestones happen every $10^3$ epochs, the initial learning rate is $10^{-2}$ and the reduction (\texttt{gamma}) factor is $0.75$. We train our model in batches of size $200$ per epoch, and each batch includes sampled probability configurations from $25$ different time steps that form the subset of the target time interval, $\tau\subset \left[0, \mathcal T\right]$. In our work we set $\mathcal T = 0.25/\Gamma$ and the time step $dt = 10^{-3}/\Gamma$ ($\mathcal T = 0.25/\Gamma_R$ and the time step $dt = 10^{-3}/\Gamma_R$ in the case of the chiral models, where $\Gamma_L < \Gamma_R$), such that the number of steps in $\left[0, \mathcal T\right]$ is equal to $250$. In Ref.~\cite{de_walle_many-body_2024} the $\tau$ subset was composed of time steps randomly picked from the interval $\left[0, \mathcal T\right]$ for every learning epoch, such that it covered all the steps homogeneously on average. In the present work, $\tau$ range is increased gradually from $0$ to $\mathcal T$ during an initial stage of training to facilitate a smoother loss reduction. Specifically, we start with the range of just first $5$ steps of the total interval, increasing it gradually each $200$ epochs by $5$, reaching the full range after $10^4$ epochs. The condition for exiting the training is either reaching the maximum number of epochs $N_\mathrm{epochs}$ or reaching the loss value that is smaller than some threshold, $\mathcal L_\theta < \epsilon_{\mathcal L}$, upon reaching the full range of $\tau$. Throughout our work the loss threshold is set to $\epsilon_{\mathcal L} = 10^{-2}$.

Upon exiting the training stage, the t-NQS output $p_\theta(\mathbf{a},t)$ is used to calculate the dynamics of physical observables appearing in Section~\ref{sec:results}. For a given physical observable and corresponding operator $O$, the expectation value can be calculated as~\cite{luo_autoregressive_2022}:
\begin{align}
    \langle O \rangle \approx \sum_{\mathbf{a}\sim p_\theta(\mathbf{a},t)} {\mathrm{tr}\left( O M_{\left(\mathbf a\right)}\right)},
\end{align}
where $M_{\left(\mathbf a\right)}$ is the IC POVM frame used to generate the probability distribution $p_\theta(\mathbf{a},t)$. At each time step of each t-NQS interval, observables are estimated using $10^3$ configurations sampled from the probability distribution, $\mathbf{a}\sim p_\theta(\mathbf{a},t)$. For all of the t-NQS results appearing in this work we additionally apply Savitzky-Golay filter to decrease the noise in the data~\cite{savitzky1964smoothing}. The Savitzky-Golay filter parameters we used in the present work: the window length is $201 dt$ and the polynomial order is $3$.

\section{Computational cost estimation}
\label{app:costs}
In this appendix we estimate the computational costs for both methods to numerically solve the ME~\eqref{eq:ME}, namely, the TN approach based on the vectorized density matrix presented in Section~\ref{sec:TN} and the t-NQS method based on IC-POVM representations of the density matrix presented in Section~\ref{sec:NQS}. It is important to note that the cost estimates for these methods should be compared with caution. Even though both of the approaches essentially rely on tensor multiplications and contractions, the basic operations behind the methods are structurally different. The TN approaches rely on a so-called singular value decomposition (SVD, see~\cite{schollwock_density_2011}) and subsequent tensor contractions. In turn, the NQS simulations imply two parts: the first one is the NN learning, where the forward and backward pass as well as the loss estimations are made, and the second one is the sampling, where the physical observables are being calculated based on the data output of the trained NN~\cite{lange_architectures_2024}. Apart from the principal differences in construction, the computer hardware on which the mentioned operations are performed is also typically different: while the NN processing uses graphics processing units (GPUs), the TN procedures still mostly rely on central processing units (CPUs), even though a significant effort is being put to transfer TN processing to GPUs~\cite{itensor}. Due to these reasons it is very challenging to draw any conclusions from comparing the absolute performance times of the TN- and NN-based methods.

We take an alternative approach to compare computational costs by defining a set of tunable parameters that influence performance times and comparing their scalings in terms of the big O notation~\cite{cormen2022introduction}. For instance, in the TN method the most computationally costly part is the SVD sweep, during which the TN that represents the state is simultaneously propagated and brought to the canonical form~\cite{schollwock_density_2011}. The cost behind such a sweep defines the general cost of the TN method and can be estimated as $\mathcal O \left(\mathcal C_\mathrm{TN}\right)$ with $\mathcal C_\mathrm{TN}$ given in Eq.~\eqref{eq:MPO_cost} of the main text. Importantly, $\mathcal C_\mathrm{TN}\left(\chi, n\right)$ is a function of the system size $n$ and the TN bond dimension $\chi$. The local Hilbert space dimension $d^2$ can also enter this estimate, however we ignore it as it is a constant (non-tunable) parameter.

The computational cost of the NQS is defined by two tunable parameters: the number of qubits $n$ and the hidden dimension of the network $d_h$. We do not include the other NN parameters like the number of layers or the batch size in our analysis due to the fact that these parameters are set constant throughout this work. Different stages of the NQS methods are influenced by the parameters $n$ and $d_h$ in a different way. {For instance, the forward pass during the learning stage scales as $\mathcal O \left(n^2 d_h + d_h^2 n\right)$~\cite{vaswani2017attention}, while the autoregressive sampling stage requires one order of $n$ more, thus scaling as $\mathcal O \left(n^3 d_h + n^2 d_h^2\right)$. The most significant computational cost comes from the estimation of the loss~\eqref{eq:loss}, where the all-to-all Lindbladian has to be explicitly applied and} which scales as $\mathcal O \left(\mathcal C_\mathrm{NQS}\right)$ with $C_\mathrm{NQS}$ given in Eq.~\eqref{eq:NQS_cost} of the main text. We note that $C_\mathrm{NQS}\left(d_h, n\right)$ is a function of the system size $n$ and the hidden dimension $d_h$.

\section{{Perturbative derivation of quadratic scaling in Fig.~\ref{fig:7}}}
\label{app:pert}

In this appendix we give a brief derivation of the quadratic scaling of the peak emission intensity $I_R^*$ with the emitter spacing parameter $\varepsilon$ (see Fig.~\ref{fig:7} and the corresponding discussion for more details). Consider the balanced ME given in Eqs.~\eqref{eq:ME} and~\eqref{eq:H_D_balanced} of the main text and assume equidistant emitter spacings, $k \delta x_{j, j+1} = 2\pi \left(1 + \varepsilon\right)$, with $\varepsilon \ll 1$ being a perturbation parameter. The output emission operator in this case can be written as:
\begin{align}
    I_R \equiv c_R^\dagger c_R / n= \frac{1}{n} \sum_{j, \ell} e^{i 2\pi \left(j - \ell\right) \varepsilon} \sigma^+_\ell \sigma^-_j.
\end{align}
One can use Taylor expansion:
\begin{align}
    &e^{i 2\pi \left(j - \ell\right) \varepsilon}\nonumber \\
    &= 1 + i 2\pi \left(j - \ell\right)\varepsilon - 2\pi^2 \left(j - \ell\right)^2 \varepsilon^2 + \dots,
\end{align}
and, assuming that the system's state at the emission peak time, $\rho^*$, is mostly defined by the permutationally symmetric sector, one can write the output emission intensity as a function of the spacing perturbation $\varepsilon$:
\begin{align}
    \langle I_R \rangle_{\rho^*} &= \frac{1}{n}\sum_{j, \ell} \langle \sigma^+_\ell \sigma^-_j \rangle_{\rho^*} +  \varepsilon\frac{i 2\pi }{n} \sum_{j \neq \ell} \langle \sigma^+_\ell \sigma^-_j \rangle_{\rho^*} \left(j - \ell\right) \nonumber \\
    & - \varepsilon^2 \frac{2\pi^2 }{n}\sum_{j \neq \ell} \langle \sigma^+_\ell \sigma^-_j \rangle_{\rho^*} \left(j - \ell\right)^2 + \mathcal{O}\left(\varepsilon^3\right),
    \label{eq:IR_exp}
\end{align}
where we used the notation $\langle O \rangle_{\rho^*} = \mathrm{tr} \left(O \rho^*\right)$. Due to the approximate permutational symmetry of ${\rho^*}$, the contributions odd in power of $\varepsilon$ are equal by modulus but opposite by sign and therefore are vanishing. If we assume that the emission intensity peaks exactly at half filling, $\sum_{j} \langle \sigma^+_j \sigma^-_j \rangle_{\rho^*} / n = 1/2$, then we can write:
\begin{align}
    I_R^* \left(\varepsilon = 0\right) \equiv \frac{1}{2} + \frac{1}{n}\sum_{j\neq\ell} \langle \sigma^+_\ell \sigma^-_j \rangle_{\rho^*}.
\end{align}
Using the numerical value of the peak output intensity in the Dicke regime with $\varepsilon = 0$,  $I_R^* \left(\varepsilon = 0\right) \approx 3.377$ (see Fig.~\ref{fig:7}), we can deduce:
\begin{align}
    \frac{1}{n} \sum_{j\neq\ell} \langle \sigma^+_\ell \sigma^-_j \rangle_{\rho^*} = I_R^* \left(0\right) - \frac{1}{2} \approx 2.877,
\end{align}
such that with the approximate permutational symmetry of ${\rho^*}$ we can write:
\begin{align}
    \frac{1}{n}\langle \sigma^+_\ell \sigma^-_j \rangle_{\rho^*} \approx \frac{2.877}{n \left(n-1\right)}~\forall j\neq \ell.
\end{align}
In this case, the quadratic term of Eq.~\eqref{eq:IR_exp} can be factorized as:
\begin{align}
     - \varepsilon^2 2\pi^2 \frac{2.877}{n \left(n-1\right)} \sum_{j \neq \ell} \left(j - \ell\right)^2.
     \label{eq:quad_1}
\end{align}
The sum appearing in Eq.~\eqref{eq:quad_1} can be simplified:
\begin{align}
    \sum_{j \neq \ell} \left(j - \ell\right)^2 = \frac{1}{6} n^2 \left(n^2 - 1\right),
\end{align}
such that the final expression for the quadratic term of Eq.~\eqref{eq:IR_exp} is:
\begin{align}
    - \varepsilon^2 \pi^2 \frac{2.877}{3} n\left(n + 1\right).
     \label{eq:quad_2}
\end{align}
This result appears as a black dotted line in Fig.~\ref{fig:7}. The quadratic fit function sketched as a white dotted line in Fig.~\ref{fig:7} is $I_R^* \left(\varepsilon\right) = f(\varepsilon \cdot 10^2)$ with $f(x) \approx -0.302 x^2 + 3.377$.

\end{document}